\documentclass[11pt,preprintnumbers,aps,amssymb,nofootinbib,amsmath,superscriptaddress,notitlepage]
{revtex4-1}
\usepackage{epsfig,epsf}
\usepackage{bm} 
\usepackage{color} 
\usepackage{slashed}
\usepackage{relsize}	
\usepackage{soul} 
\usepackage{hyperref}
\newcommand{\beq}{\begin{equation}}
\newcommand{\beql}[1]{\begin{equation}\label{#1}}
\newcommand{\eeq}{\end{equation}}
\def\bal#1\gal{\begin{align}#1\end{align}}
\newcommand{\ball}[1]{\bal\label{#1}}
\newcommand{\eq}[1]{(\ref{#1})}
\newcommand{\fig}[1]{Fig.~\ref{#1}}
\renewcommand{\sec}[1]{Sec.~\ref{#1}}
\renewcommand{\b}[1]{{\bm #1}} 

\newcommand{\e}{\varepsilon}
\newcommand{\aver}[1]{\left\langle #1 \right\rangle}

\begin{document}

\title{Photon radiation in hot nuclear matter by means of chiral anomalies}

\author{Kirill Tuchin}

\affiliation{
Department of Physics and Astronomy, Iowa State University, Ames, Iowa, 50011, USA}

\date{\today}

\begin{abstract}

A new mechanism of photon emission in the quark-gluon plasma is proposed. Photon dispersion relation in the presence of the $CP$-odd topological regions generated by the chiral anomaly acquires an imaginary mass. It allows photon radiation through the decay $q\to q\gamma$ and annihilation  $q\bar q\to \gamma$ processes closely related to the chiral Cherenkov radiation. Unlike previous proposals this mechanism does not require an external magnetic field. The differential photon emission rate per unit volume is computed and shown to be comparable to the rate of photon emission in conventional  processes. 

\end{abstract}

\maketitle

\section{Introduction}\label{sec:i}

Photon radiation by hot nuclear matter has been a focus of experimental and theoretical studies for many decades. 
However, in spite of considerable progress, there are still unresolved problems concerning the photon spectrum produced in relativistic heavy ion collisions such as the puzzling enhancement of the direct photon production \cite{Adare:2008ab,Adam:2015lda}. The major contributors to the photon spectrum are the quark--antiquark annihilation  and the QCD Compton scattering processes in the quark-gluon plasma and the inelastic reactions in the hot hadronic gas \cite{Baier:1991em,Kapusta:1991qp,Hung:1996mq,Steele:1996su,Steele:1997tv,Dusling:2009ej,Lee:1998nz,Aurenche:2000gf, Arnold:2001ms, Peitzmann:2001mz,Turbide:2003si,Bratkovskaya:2008iq,Vitev:2008vk,vanHees:2011vb,Paquet:2015lta,Linnyk:2015tha}. In addition to these ``conventional" processes, other contributors have been proposed such as photon emission by the nuclear matter before the QGP formation  \cite{Chiu:2012ij,McLerran:2014hza}, the synchrotron radiation  \cite{Tuchin:2012mf,Yee:2013qma}, radiation via the conformal anomaly  \cite{Basar:2012bp} and through the chiral anomaly \cite{Fukushima:2012fg} as well as the modification of the conventional processes due to the axial charge fluctuations \cite{Mamo:2013jda,Mamo:2015xkw}.  The mechanisms suggested in  Refs.~\cite{Tuchin:2012mf,Basar:2012bp,Fukushima:2012fg} rely on existence of intense magnetic field produced in heavy-ion collisions. In this paper we argue that there is a different unconventional mechanism of photon production via the chiral anomalies of QED and QCD which \emph{does not} involve the external magnetic field.

The hot nuclear matter, or quark-gluon plasma (QGP), is believed to contain the topological $CP$-odd domains created by the random sphaleron-mediated  transitions between different QCD vacua. Interaction of the electromagnetic field with these domains can be described by adding to the QED Lagrangian the  axion-photon coupling term  \cite{Fujikawa:2004cx} 
\ball{i1}
\mathcal{L}_A =-\frac{c_A}{4}\theta F_{\mu\nu}\tilde F^{\mu\nu}\,,
\gal
where $c_A= N_c\sum_f q_f^2e^2/2\pi^2$ is the QED anomaly coefficient and the field $\theta$ is sourced by the topological charge density
\ball{i2}
q(x)= \frac{g^2}{32\pi^2}G_{\mu\nu}^a\tilde G^{a\mu\nu}(x) 
\gal
which varies in space and time across a $CP$-odd domain. As a result \eq{i1} cannot be rewritten as a total derivative and removed from the Lagrangian. Instead, it appears in the modified Maxwell equations as the spatial and the temporal derivatives of $\theta$.

It has been known since the pioneering article by Carroll, Feld and Jackiw \cite{Carroll:1989vb} that in QED coupled to the axion field, photons acquire an imaginary mass $m_A$ making possible  their  spontaneous emission by fermions.  This phenomenon is referred to as the vacuum Cherenkov radiation \cite{Lehnert:2004hq,Lehnert:2004be}.
Since the electromagnetic field in QGP is coupled to the axion field $\theta$, it is natural to expect that a similar mechanism of photon radiation exists in hot nuclear medium as well.  This idea was developed in  \cite{Tuchin:2018sqe,Huang:2018hgk} where it was argued that ultrarelativistic fermions moving in a finite-$\theta$ domain radiate photons, which we referred to as  the \emph{chiral} Cherenkov radiation. Additionally, fermions in QGP radiate the \emph{chiral} transition radiation  as they cross the boundary between the plasma and vacuum due to the difference in the photon wave function inside and outside the plasma. The spectra of both processes are  proportional to the average values of the spatial and  the temporal $\theta$-derivatives. Since the chiral Cherenkov radiation scales with the system volume, whereas the chiral transition radiation scales with its area, the former is dominant when the contribution of the entire QGP (as opposed to a single fast quark) is considered. Thus, the present work focuses on the chiral Cherenkov radiation by QGP. 
 
The analysis of  \cite{Tuchin:2018sqe,Huang:2018hgk} relied on two basic assumptions: (i) $\theta(x)$ is a slowly varying adiabatic function of its arguments and (ii) the absolute value of the photon mass generated by the anomaly $|m_A|$ is much larger than the plasma frequency $\omega_\text{pl}$. The first assumption is the simplest model that captures the essential dynamics of the chiral magnetic effect \cite{Kharzeev:2004ey,Kharzeev:2007tn,Fukushima:2008xe,Kharzeev:2009fn}. It is supported by the results obtained by Zhitnitsky \cite{Zhitnitsky:2014ria,Zhitnitsky:2014dra}. The second assumption is justified for large enough photon energy $\omega$ because $m_A^2$ is proportional to $\omega$, see \eq{b5},  whereas the plasma frequency is $\omega$-independent. These are the assumptions that are carried over to the present study as well. However, unlike the radiation by a single quark discussed in  \cite{Tuchin:2018sqe,Huang:2018hgk} where one is free to choose the quark energy high enough so that most of the photon spectrum satisfy $|m_A|\gg \omega_\text{pl}$, in the case of QGP the bulk of the photon radiation occurs at $\omega\lesssim T$, where $T$ is the QGP temperature. Still, it is argued in the next section that 
at high enough temperatures, the photon mass satisfies the second assumption since the plasma frequency is  proportional to $T$, see \eq{b2}, whereas the absolute value of $m^2_A$ is proportional to the sphaleron transition rate $\Gamma$ which rises at high temperatures as $T^4$. 

The paper is organized as follows.  \sec{sec:a} deals with the qualitative discussion of the electromagnetic fields in presence of the $CP$-odd domains. The mean value of the $\theta$-field in a domain is related to the sphaleron transition rate and hence  scales as $T^4$ at high temperatures. This indicates that at high enough temperatures the chiral Cherenkov radiation becomes possible. In \sec{sec:b}, the photon dispersion relation at finite $\theta$ is discussed. The main section is \sec{sec:d} where the photon radiation rate is computed. 
In order to simplify the derivations and emphasize the main physics points, I am going to consider the relativistic limit $\omega\gg |m_A|$; generalization beyond this limit is straightforward. In fact, such a generalization for a single quark has been recently obtained in \cite{Tuchin:2018mte}. The discussion and summary is presented in \sec{sec:f}.


\section{Electrodynamics in quark-gluon plasma  with $CP$-odd domains}\label{sec:a}

The $CP$-odd domains in the chiral matter can be  described by the axion -- a pseudo-scalar field $\theta$ whose interactions with the electromagnetic  $F_{\mu\nu}$ and color $G^a_{\mu\nu}$ fields are governed by the Lagrangian \cite{Wilczek:1987mv,Carroll:1989vb,Sikivie:1984yz,Kalaydzhyan:2012ut}
\ball{a1}
\mathcal{L}=\mathcal{L}_\text{QED}+\mathcal{L}_\text{QCD}  -\frac{c_A}{4} \theta   F_{\mu\nu}\tilde F^{\mu\nu}-\frac{c_A'}{4} \theta   G^a_{\mu\nu}\tilde G^{a\mu\nu}  + f^{2}\left[\frac{1}{2}(\partial_\mu \theta)^2-  \frac{1}{2}m^2_\text{ax} \theta^2\right]\,,
\gal
where $\tilde F_{\mu\nu}= \frac{1}{2}\epsilon_{\mu\nu\lambda\rho} F^{\lambda\rho}$ is the dual field tensor,  $c_A$,  $c_A'$ are the QED and QCD anomaly coefficients respectively  and $f$, $m_\text{ax}$ are  constants with mass dimension one. It follows that the axion equation of motion is 
\ball{a2}
(\partial^2+m_\text{ax}^2)\theta = -\frac{1}{4f^2}\left( c_A' G_{\mu\nu}^a\tilde G^{a\mu\nu} +c_A F_{\mu\nu}\tilde F^{\mu\nu}\right)
\,.
\gal
In the quark-gluon plasma the electromagnetic contribution to the topological charge density is presumed to be negligible so that the axion dynamics is driven primarily by the topologically non-trivial gluon configurations. Assuming  that $\theta$ is slowly varying inside a $CP$-odd domain one can express it in terms of the  topological number density \eq{i2} as
\ball{a4}
\theta(x)= -\frac{N_fq(x)}{f^2m_\text{ax}^2}\,.
\gal

The equations of motion  of electromagnetic field coupled to the axion field read
\bal
&\partial_\mu F^{\mu\nu}= j^\nu- c_A \tilde F^{\mu\nu}\partial_\mu\theta\,,\label{a5}\\
&\partial_\mu \tilde F^{\mu\nu}= 0\,.\label{a6}
\gal
%
In a slowly varying field $\theta$, its first derivatives $\partial^\mu\theta$ can be replaced by their constant domain--average values denoted by $\sigma_\chi= c_A\dot \theta$\cite{Fukushima:2008xe,Kharzeev:2009fn,Kharzeev:2009pj}, referred to as the chiral conductivity, and $ \b b=c_A\b\nabla \theta$. In this approximation photon and axion dynamics decouple and one can consider electrodynamics in the topologically non-trivial background \cite{Kharzeev:2013ffa}.

The average of the axion field over an ensemble of $CP$-odd domains vanishes. However, its value in a single domain can be finite due to the fluctuations of the topological number $N_{CS}$. In the context of this work we need to know the temperature dependence of the axion field in a domain because it determines the temperature dependence of the effective photon mass $m_A$. In particular, if its $T$-dependence is steeper than linear, then one expects that there is a range of temperatures where the plasma becomes radioactive as explained at the end of  \sec{sec:i}. The topological number density can be estimated as $q\sim N_{CS}/V_\text{dom}$, where $V_\text{dom} \sim 1/m_\text{ax}^4$ is the domain 4-volume. Since the sphaleron size is inversely proportional to $T$, the domain volume decreases as  $V_\text{dom}\sim 1/T^4$.  Fluctuations of $N_{CS}$ are related to the sphaleron transition rate $\Gamma$ as  $\aver{N_{CS}^2}= 2\Gamma V_\text{pl}$ \cite{Rubakov:1996vz} for large enough  4-volume $V_\text{pl}$ of plasma. Therefore, the variance of the topological number density is $\aver{q^2}\sim m_\text{ax}^8\Gamma V_\text{pl}$. Employing \eq{a4}  it is seen that  the typical variance of the axion field strength is $\aver{\theta^2} \sim m_\text{ax}^4  \Gamma V_\text{pl}/f^4$.
$\Gamma$ is exponentially suppressed at low temperatures, but increases as $T^4$ at high temperatures \cite{Moore:1997sn,Bodeker:1999gx,Moore:2010jd,Son:2002sd}. It follows, using \eq{b5} of the next section,  that $m_A\sim \aver{\theta}^{1/2}\sim T^4$. Thus, $|m_A|$ exceeds $\omega_\text{pl}$ at high $T$ making the chiral Cherenkov radiation possible.

\section{Photon dispersion relation}\label{sec:b}

Now that the model parameters  have been outlined, it is instructive to review  the photon dispersion relation. In the  case $\theta=0$ the photon dispersion relation at finite temperature $T$ and finite chemical potentials  of the right and left-handed fermions $\mu_{R,L}$ was computed  in \cite{Akamatsu:2013pjd}. In the high-energy limit, when the photon is near the mass-shell and transversely polarized, its dispersion relation is $\omega^2-k^2= \omega^2_\text{pl}$, where 
\ball{b2}
\omega^2_\text{pl}=\frac{m_D^2}{2} &= \frac{e^2}{2}\left( \frac{T^2}{6}+ \frac{\mu^2}{2\pi^2}\right)\,, 
\gal
and $\mu^2= \mu_R^2+\mu_L^2$. 

At finite $\theta$ the photon dispersion relation acquires an extra term due its interaction with the $CP$-odd domains 
\ball{b4}
\omega^2-k^2= \omega^2_\text{pl} + m_A^2+ \mathcal{O}(\omega-k)\,,
\gal
where $m_A^2$ is given by 
\ball{b5}
m_A^2= -\lambda \sigma_\chi \omega\,,\qquad \text {or}\qquad m_A^2= -\lambda \b k\cdot \b b\,,
\gal
depending on which of the parameters $\sigma_\chi$ or $b$ is largest \cite{Huang:2018hgk}\footnote{In \cite{Huang:2018hgk} $m_A$  was denoted as $\mu$. The dispersion relations for arbitrary $\sigma_\chi$ and $b$ can be found in \cite{Carroll:1989vb}.} and  $\lambda= \pm 1$ is the right and left-handed photon polarization. Note that $m_A$ can be real or imaginary. As explained in the previous two sections, at high enough photon energies and plasma temperatures $\omega_\text{pl}$ is but a small correction compared to $m_A$ and will be neglected in the following sections.

\section{Photon radiation rate}\label{sec:d}

Photon emission by means of the chiral Cherenkov radiation mechanism can proceed via two channels: (i) the decay channel  $q\to q\gamma$ and (ii) the annihilation channel $q \bar q\to  \gamma$.\footnote{I am using the term `the chiral Cherenkov radiation' with respect to both channels.} The total photon radiation rate is the sum of rates of these two processes.  

\subsection{Decay channel}

The scattering matrix element for photon radiation in the decay channel $q(p)\to q(p')+\gamma(k)$ is given by
$S_D=(2\pi)^4\delta^{(4)}(p'+k-p) i\mathcal {M}_D$ where 
\ball{c5}
i\mathcal {M}_D
=-ie Q \frac{\bar u_{\b p' s'}\slashed{\epsilon}^*_{\b k \lambda} u_{\b p s}}{\sqrt{8\e\e' \omega V^3}}\,.
\gal
The components of the 4-vectors are $p= (\e, \b p)$, $p'= (\e', \b p')$ and $k=(\omega, \b k)$, $Q$ is quark charge and $m= gT/\sqrt{3}$ its thermal mass \cite{Arnold:2001ms}. We retained the relativistic normalization factors $(2p^0 V)^{-1/2}$ for each of the three fields, where $V$ is the normalization volume. The radiation probability can be computed as
\ball{c8}
dw_D= 2N_c \frac{1}{2}\sum_{\lambda s s'}|S_D|^2 f(\e)[1-f(\e')] \frac{V d^3p'}{(2\pi)^3}\frac{V d^3k}{(2\pi)^3}\frac{V d^3p}{(2\pi)^3}\,,
\gal
where $2N_c$ accounts for the number quarks and antiquarks of different color, $1/2$ comes from the incident quark spin average  and $f(\e)$ is the quark equilibrium distribution function, which reads 
\ball{c8.1}
f(\e)= \frac{1}{e^{\e/T}+1}\,.
\gal
The small chemical potentials of quarks is neglected.  
The rate of photon production per unit volume can be computed as
\ball{c9}
d\Gamma_D= \frac{dw_D}{VT}= 2N_c\frac{\delta(\omega+\e'-\e)}{16 (2\pi)^5\e\e'\omega}\sum_{\lambda s s'}|i\mathcal {M}_D|^2f(\e)[1-f(\e')]d^3k d^3p\,.
\gal
Performing the summation over the transverse photon polarizations  using 
\ball{c10}
\sum_\lambda \epsilon_{\b k\lambda}^\mu \epsilon_{\b k\lambda}^{\nu*}= \left\{\begin{array}{cc}0, & \nu \mu =0\,, \\\delta^{ij}-\frac{k^ik^j}{k^2}, & \nu=i, \mu = j\,. \end{array}\right.
\gal
 yields the result
\ball{c10.1}
 \sum_{ss'}|\mathcal{M}_D|^2=4\left[\e\e'-m^2- \frac{(\b k\cdot \b p)(\b k\cdot \b p')}{\b k^2}\right]\,.
\gal
In the high energy limit the momenta of the initial and final quarks and the photon have a large component, say along the $z$-direction, that allows one to  approximate
\ball{c11}
p_z&\approx \e \left( 1- \frac{p_\bot^2+m^2}{2\e^2}\right)\,,\quad k_z\approx \omega \left( 1- \frac{k_\bot^2+m_A^2}{2\omega^2}\right)\,,\quad p_z'\approx \e'\left( 1- \frac{p'^2_\bot+m^2}{2\e'^2}\right)\,.
\gal
Denoting by  $x= \omega/\e$ the fraction of the incident quark energy carried away by the photon and substituting \eq{c11} into \eq{c10}  one derives 
\ball{c12}
\sum_{ss'}|\mathcal{M}_D|^2=\frac{2}{x^2(1-x)}\left[q_\bot^2(2-2x+x^2)+m^2x^4\right]\,,
\gal
where $\b q_\bot= x\b p_\bot-\b k_\bot$. In the same approximation the energy delta-function can be written as 
\ball{c14}
\delta(\omega+\e'-\e)\approx 2x(1-x)\e \delta\left(q_\bot^2+m_A^2(1-x)+m^2x^2\right)\,,
\gal
Substituting \eq{c12} and \eq{c14} into \eq{c9} and integrating over $q_\bot$ instead of $p_\bot$ one finds 
\ball{c15}
\omega\frac{d\Gamma_D}{d^3k}
= 2N_c\frac{e^2Q^2\pi }{4(2\pi)^5 }\int_0^1  \frac{dx}{x^4}f\left(\frac{\omega}{x}\right)\left[1-f\left(\frac{\omega(1-x)}{x}\right)\right]\sum_\lambda \left\{ -m_A^2[(1-x)^2+1]-2m^2x^2\right\} \theta(-\kappa_\lambda)\,,
\gal
where it is denoted 
\ball{c16}
\kappa_\lambda = m_A^2(1-x)+m^2x^2\,.
\gal
Evidently, since $m^2>0$ the non-vanishing contribution to the photon production rate in this channel exists only if $m_A^2 <0$. Moreover, $\kappa_\lambda$ is negative only  if $|m_A^2|(1-x)>x^2m^2$ which occurs when
\ball{c17}
0\le x < \frac{|m_A^2|}{2m^2}\left(\sqrt{1+\frac{4m^2}{|m_A^2|}}-1\right)\,.
\gal

One can perform the integration of $x$ explicitly in the limit $m\ll |m_A|$. It is convenient to introduce a new variable  $\xi = 1/x-1$ in place of $x$ and rewrite \eq{c15} as 
\ball{c20}
\omega\frac{d\Gamma_D}{d^3k}=2N_c\frac{e^2Q^2}{8(2\pi)^4 }\int_0^\infty d\xi \left\{ -m_A^2\left[ \xi^2+(1-\xi)^2\right]-2m^2\right\} f(\omega(1+\xi))\left[ 1-f(\omega\xi)\right]&\nonumber\\
\times\theta\left( -m_A^2\xi(1+\xi)-m^2\right)\,, &
\gal
where only the photon polarization that gives $m_A^2<0$ contributes. Neglecting $m$ one obtains
\ball{c21}
\omega\frac{d\Gamma_D}{d^3k}=2N_c\frac{ e^2Q^2}{8(2\pi)^4 } |m_A^2|\int_0^\infty d\xi \left[ \xi^2+(1-\xi)^2\right]f(\omega(1+\xi))\left[ 1-f(\omega\xi)\right]\,.
\gal
Note that the condition \eq{c17} is now trivial  $0<x<1$. Also, since $e\ll g$, $\omega_\text{pl}\ll m$ implying that $m_A\approx m_A$ in this approximation. We also approximate $1-f(\omega\xi)\approx 1-(e+1)^{-1}=0.73$ since the argument of $f$ is typically on the order of unity, for otherwise the distribution $f(\omega(1+\xi))$ of the  incident quark is exponentially suppressed. Thus we derive
\ball{c22}
\omega\frac{d\Gamma_D}{d^3k}=0.73\cdot 2N_c\frac{e^2Q^2}{8(2\pi)^4 } |m_A^2|\left[ \frac{\ln(1+e^{-\beta\omega})}{\beta \omega}+\frac{2\text{Li}_2(-e^{-\beta\omega})}{(\beta\omega)^2}-\frac{4\text{Li}_3(-e^{-\beta\omega})}{(\beta\omega)^3}\right]\,.
\gal
The low and high energy regions of  the spectrum read 
\ball{c23}
\omega\frac{d\Gamma_D}{d^3k}=0.73\cdot 2N_c\frac{e^2Q^2}{8(2\pi)^4 } |m_A^2|
\left\{\begin{array}{ll}\frac{3\zeta(3)}{(\beta \omega)^3}\,, &  \omega\ll T \\
 \frac{1}{\beta\omega}e^{-\beta\omega}\,,&  \omega\gg T\,.\end{array}\right.
\gal
Taking into account that $m_A^2$ is proportional to $\omega$, we find that at $\omega\ll T$, the photon  of spectrum scales as $1/\omega^2$. Thus the total photon rate $\Gamma_D$ is dominated by soft photons $\omega\ll T$ that produce the large logarithm $\ln (T/m)$.

\subsection{Annihilation channel}

The scattering matrix element for photon radiation in the annihilation channel $q(p)+\bar q(p_1)\to \gamma(k)$ is given by $S_A=(2\pi)^4\delta^{(4)}(p+p_1-k) i\mathcal {M}_A$ where 
\ball{d5}
i\mathcal {M}_A
=-ie Q \frac{\bar v_{\b p_1 s_1}\slashed{\epsilon}^*_{\b k \lambda} u_{\b p s}}{\sqrt{8\e\e_1 \omega V^3}}\,.
\gal
The corresponding radiation probability can be computed as
\ball{d8}
dw_A= N_c\frac{1}{4}\sum_{\lambda s s'}|S_A|^2 f(\e)f(\e_1) \frac{V d^3p_1}{(2\pi)^3}\frac{V d^3k}{(2\pi)^3}\frac{V d^3p}{(2\pi)^3}\,
\gal
where $N_c$ accounts for different colors and $1/4$ stems from the incident quark and antiquark spin average. The rate of photon production per unit volume reads
\ball{d9}
d\Gamma_A= \frac{dw_A}{VT}= N_c\frac{\delta(\omega-\e_1-\e)}{32 (2\pi)^5\e\e_1\omega}\sum_{\lambda s s_1}|i\mathcal {M}_A|^2f(\e)f(\e_1)d^3k d^3p\,.
\gal
Summation over the transverse photon polarizations using \eq{c10} yields  
\ball{d10}
 \sum_{ss_1}|\mathcal{M}_A|^2=4\left[\e\e_1+m^2- \frac{(\b k\cdot \b p)(\b k\cdot \b p_1)}{\b k^2}\right]\,.
\gal
Employing the high energy limit \eq{c11} and denoting by  $y= \e/\omega$ the energy fraction that the incident quark  contributed to the photon energy and $\b \ell_\bot= y\b k_\bot-\b p_\bot$ one derives 
\ball{d12}
\sum_{ss_1}|\mathcal{M}_A|^2=\frac{2}{y(1-y)}\left[\ell_\bot^2\left(y^2+(1-y)^2\right)+m^2\right]\,,
\gal
and 
\ball{d14}
\delta(\omega-\e_1-\e)\approx 2y(1-y)\omega \delta\left(\ell_\bot^2-m_A^2y(1-y)+m^2\right)\,.
\gal
These formulas can also be obtained from the results of the previous subsection using the crossing-symmetry.
Substituting \eq{d12} and \eq{d14} into \eq{d9} and integrating over $\ell_\bot$ instead of $p_\bot$ one finds 
\ball{d15}
\omega\frac{d\Gamma_A}{d^3k}
= N_c\frac{e^2Q^2\pi }{8(2\pi)^5 }\int_0^1  dy f(y\omega)f((1-y)\omega)\sum_\lambda \left[ m_A^2(2y^2-2y+1)+2m^2 \right] \theta(- \varkappa_\lambda)\,,
\gal
where it is denoted 
\ball{d16}
 \varkappa_\lambda = -m_A^2y(1-y)+m^2\,.
\gal
In the annihilation channel $m_A^2$ must be positive in order that $\varkappa_\lambda$ be negative. Additionally, the energy fraction $y$ is restricted to the interval
\ball{d20}
\frac{1}{2}\left(1-\sqrt{1-\frac{4m^2}{|m_A^2|}}\right)< y < \frac{1}{2}\left(1+\sqrt{1-\frac{4m^2}{|m_A^2|}}\right)\,.
\gal
Clearly, the radiation is possible only if $|m_A|>2m$.

\begin{figure}[ht]
      \includegraphics[height=5cm]{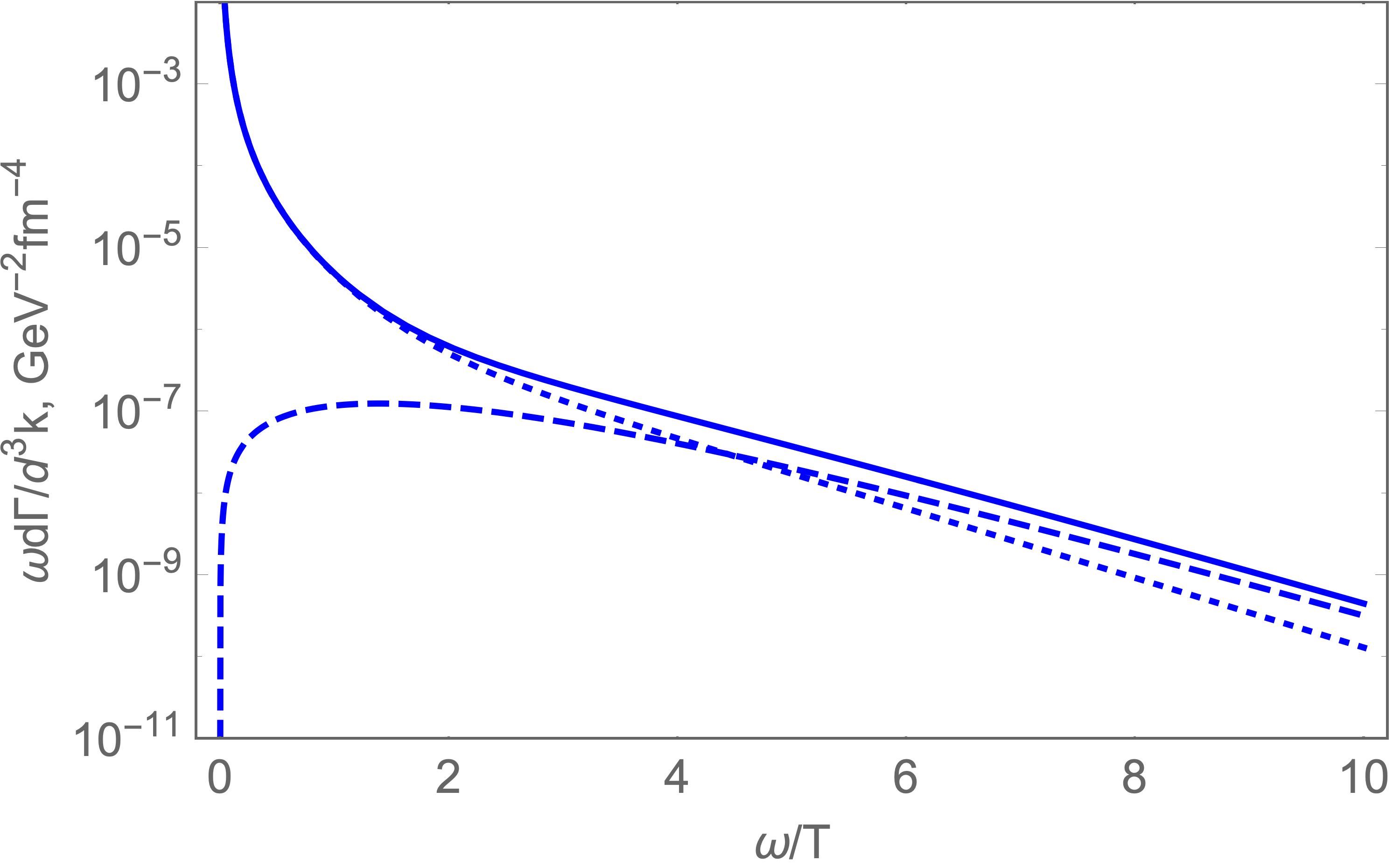} 
  \caption{Differential photon emission rate (solid line) and its two contributions from the decay (dashed line) and annihilation (dotted line) processes.  Plasma temperature $T=400$~MeV, chiral conductivity $\sigma_\chi = 1$~MeV  and $\sum_f Q_f^2 = 5/9$ (for the two lightest flavors). }
\label{fig1}
\end{figure}

In the limit $|m_A|\gg m$, \eq{d15} simplifies 
\ball{d22}
\omega\frac{d\Gamma_A}{d^3k}= N_c\frac{e^2Q^2 }{16(2\pi)^4 }|m_A^2|\int_0^1  dy f(y\omega)f((1-y)\omega)(2y^2-2y+1)\,,
\gal
where only the photon polarization that gives $m_A^2>0$ contributes. The integral can be taken exactly:
\ball{d33}
&\int_0^1  dy f(y\omega)f((1-y)\omega)(2y^2-2y+1)\nonumber\\
=&
\frac{1}{e^{\beta\omega}-1}\left[-\frac{8\text{Li}_3(-e^{\beta\omega})+6\zeta(3)}{(\beta\omega)^3}
+\frac{4\text{Li}_2(-e^{\beta\omega})-\pi^2/3}{(\beta\omega)^2}
+\frac{2\ln(1+e^{\beta\omega})- \ln 4}{\beta\omega}-\frac{2}{3}\right]\,.
\gal
At low and high photon energy the spectrum reads 
\ball{d23}
\omega\frac{d\Gamma_A}{d^3k}=N_c\frac{e^2Q^2}{16(2\pi)^4 } |m_A^2|
\left\{\begin{array}{ll}\frac{1}{6}\,, &  \omega\ll T \\
 \frac{2}{3}e^{-\beta\omega}\,,&  \omega\gg T\,.\end{array}\right.
\gal
Comparing with \eq{c23} one can see that the decay channel dominates the low energy part of the spectrum, whereas the annihilation channel dominates the high energy tail, see \fig{fig1}.  It is remarkable that since the photon polarization in the two channels is opposite, the total photon spectrum has different polarization direction at low and high energies with respect to $T$. 

\section{Discussion and summary}\label{sec:f}

The main result of this paper are Eqs.~\eq{c15} and \eq{d15} that represent the differential rates of  photon emission rate by means of the chiral Cherenkov radiation in the decay and annihilation channels. Their sum gives the total photon emission rate (per unit volume). The magnitude of this contribution to the total photon yield by QGP shown in \fig{fig1}  is comparable with the conventional contributions as one can glean from  Fig.~3 of \cite{Paquet:2015lta}.

An important phenomenological question  is the value of the photon emission threshold at a given QGP temperature. Electromagnetic radiation by means of the mechanism described in this paper is possible if  $\omega_\text{pl}< |m_A|$.  The plasma frequency \eq{b2} of QGP at temperature $T=400$~MeV is 
$\omega_\text{pl}\approx 35$~MeV. The chiral conductivity is unknown, but is often estimated to be of the order of a fraction of MeV. Importantly,  it rapidly increases as $T^4$. Thus, for example, if $\sigma_\chi = 1-10$~MeV then using the first of the equations \eq{b5},  the infrared photon emission threshold is $\omega_0\sim 0.1-1$~GeV. This is certainly within the range of phenomenologically interesting photon energies. A more precise knowledge of $\sigma_\chi$ may be extracted from the measurements of the charge separation effect in relativistic heavy-ion collisions because this effect is generated by the anomalous electric current proportional to $\sigma_\chi$ \cite{Kharzeev:2007tn}.

The calculations performed in this paper relied on the high energy approximation in which $\omega\gg m$. Since the quark thermal mass $m$ is of the order of a hundred MeV, this approximation is not sufficiently reliable for the phenomenological applications to the QGP at realistic temperatures. Still, considering this limit has a great advantage of emphasizing the physics mechanism of photon radiation with the least mathematical and numerical complications possible. A comprehensive phenomenological approach would of course require going beyond the high-energy approximation. 

\acknowledgments
This work was supported in part by the U.S. Department of Energy under Grant No.\ DE-FG02-87ER40371.


\end{document}